 \documentclass[final,5p,times,twocolumn]{elsarticle}

\usepackage{graphicx}

\usepackage{amssymb}

\journal{Nuclear Instruments and Methods B}
\begin{document}
\begin{frontmatter}
\title{A Novel Ion Cooling Trap for Multi-Reflection Time-of-Flight Mass Spectrograph}
\author[1,2]{Y.~Ito}
\author[1,2,6]{P.~Schury}
\author[1]{M.~Wada}
\author[1]{S.~Naimi}
\author[1,3]{C.~Smorra}
\author[1]{T.~Sonoda}
\author[1,2]{H.~Mita}
\author[1,4]{A.~Takamine}
\author[1,5]{K.~Okada}
\author[2]{A.~Ozawa}
\author[1,6]{H.~Wollnik}
\address[1]{SLOWRI Team, Nishina Accelerator-Based Research Center, RIKEN, 2-1 Hirosawa, Wako, Saitama 351-0198, Japan}
\address[2]{University of Tsukuba, 1-1-1 Tennodai, Tsukuba, Ibaraki 305-8577, Japan}
\address[3]{Max-Planck-Institut f\"{u}r Kernphysik, Saupfercheckweg 1, D-69117 Heidelberg, Germany}
\address[4]{Aoyama Gakuin University, 4-4-25 Shibuya, Shibuya-ku, Tokyo 150-8366, Japan}
\address[5]{Sophia University, 7-1 Kioi-cho, Chiyoda-ku, Tokyo 102-8554, Japan}
\address[6]{New Mexico State University, Department Chemistry and Biochemistry, Las Cruces, NM 88003,USA}
\begin{abstract}
A radiofrequency quadrupole ion trap system for use with a multi-reflection time-of-flight mass spectrograph (MRTOF) for short-lived nuclei has been developed.
The trap system consists of two different parts, an asymmetric taper trap and a flat trap.
The ions are cooled to a sufficient small bunch for precise mass measurement with MRTOF in only 2 ms cooling time in the flat trap, then orthogonally ejected to the MRTOF for mass analysis.
A trapping efficiency of $\approx$27$\%$ for $^{23}$Na$^{+}$ and $\approx$5.1$\%$ for $^{7}$Li$^{+}$ has been achieved.  
\end{abstract}
\end{frontmatter}

\section{Introduction}
\label{}
\par Precision mass measurements of exotic nuclei with a multi-reflection time-of-flight mass spectrograph (MRTOF) will be key experiments at the SLOWRI facility at RIKEN \cite{Wollnik1990, Wada2003, Ishida2005, Schury2009}.
The MRTOF is expected to be one of the most powerful mass measurement devices, competitive with standard Penning trap mass spectrometer (PTMS) \cite{Schury2013}.
In the experiments for exotic nuclei, the advantages of the MRTOF over the standard PTMS are shorter measurement times for medium and heavy masses, and reduced yield requirements as all detected ions contribute equally to the statistics.
These features will allow us to measure, for instance, masses related to r-process nucleosynthesis \cite{Schatz2006} as well as superheavy elements with short life-times ($T_{1/2}\lesssim100$~ms) and low production yields.
\par The MRTOF is coupled with a gas cell to decelerate and thermalize high-energy radioactive ions; a multipole ion guide system transports the extracted low-energy ion beam to the MRTOF \cite{Takamine2005}.
As MRTOF mass measurements are based on time-of-flight (ToF) measurement, the continuous beam extracted from the gas cell must be converted into a pulsed beam.
An optimal time-of-flight measurement would feature an extremely long flight path traversed by ions with a short-duration time-structure and well-defined energy.
For this purpose, a novel ion cooling trap system, specialized for rapid cooling and creation of ion pulses with excellent optical properties, has been developed.
We have characterized the trapping efficiency of the trap system with several figures of merit and found it to be very well-suited for the MRTOF.
\section{Trap system}
\label{}
\par In order to convert the continuous beam delivered from the gas cell, a buffer gas filled ion trap must be used.
The use of gas-filled radiofrequency multipole ion traps to accumulate and cool ions is a proven technique \cite{Herfurth2001}.
Our trap system is located at the end of the transport line from the gas cell, just prior to the MRTOF.
It consists of a ``taper trap" and a novel ``flat trap".
Owing to the flat trap geometry, ions can easily be accumulated from both directions and ejected orthogonally.
\par
\begin{figure}[htbp]
 \centering
 \includegraphics[bb=16 31 998 660, width=86mm]{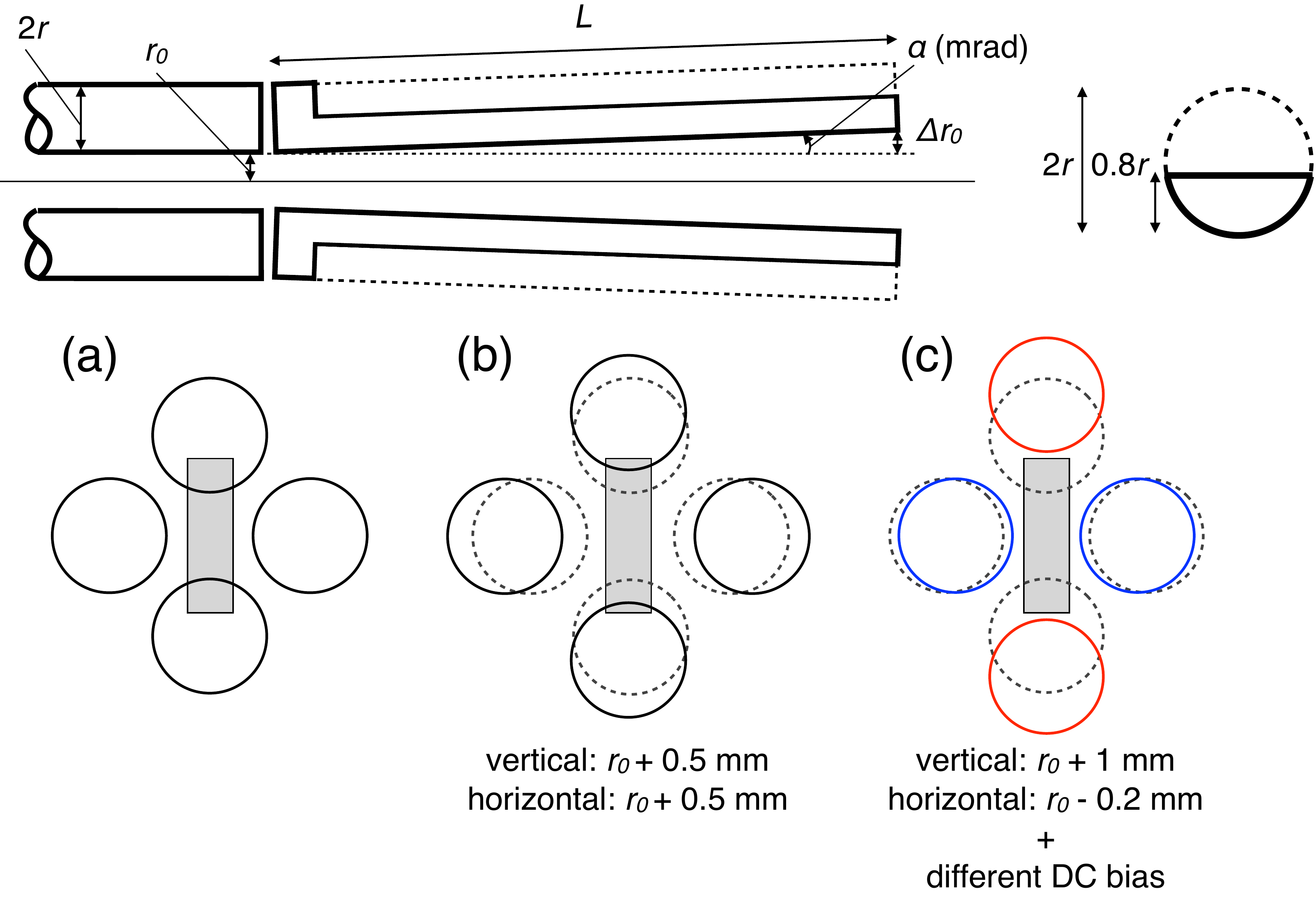}  
 \caption{(color online).
 (top) Schematic view of the taper trap.
 In order to fit into the stainless tube, the rods are 4~mm thick circular segments.
 (bottom) Cross sectional view of the rods in three configurations: (a) the parallel trap, (b) the taper trap and (c) the asymmetric taper trap (see text).
 The cross sectional view of entrance side is indicated by dashed line and the flat trap aperture is indicated by rectangular gray area in bottom figures.
 The rods are mounted in PEEK blocks at each end.
 }
 \label{fig:taper_trap_geom_001}
\end{figure}

 The taper trap illustrated in Fig. \ref{fig:taper_trap_geom_001} consists of four tilted rods with radius of $r~=~5$~mm and length of $L = 190$~mm.
The interrod-gap radius $r_{0}$ at the outer end is 4.35~mm, while at the inner end it is slightly larger, $r_{0} + \Delta r_{0}$, due to the tapered structure.
The optimal $\Delta r_{0}$ was chosen based on SIMION \cite{SIMION8} simulations.
This tapered structure can produce an effective axial drag force \cite{Mansoori1998} without the usual need for segmentation of the rods.
The taper trap is housed in a stainless steel tube to isolate it from the vacuum region outside the flat trap; a collimator at the entrance reduces the gas flow into the vacuum region.
\par
\begin{figure}[htbp]
 \centering
 \includegraphics[bb=2 140 896 800, width=86mm]{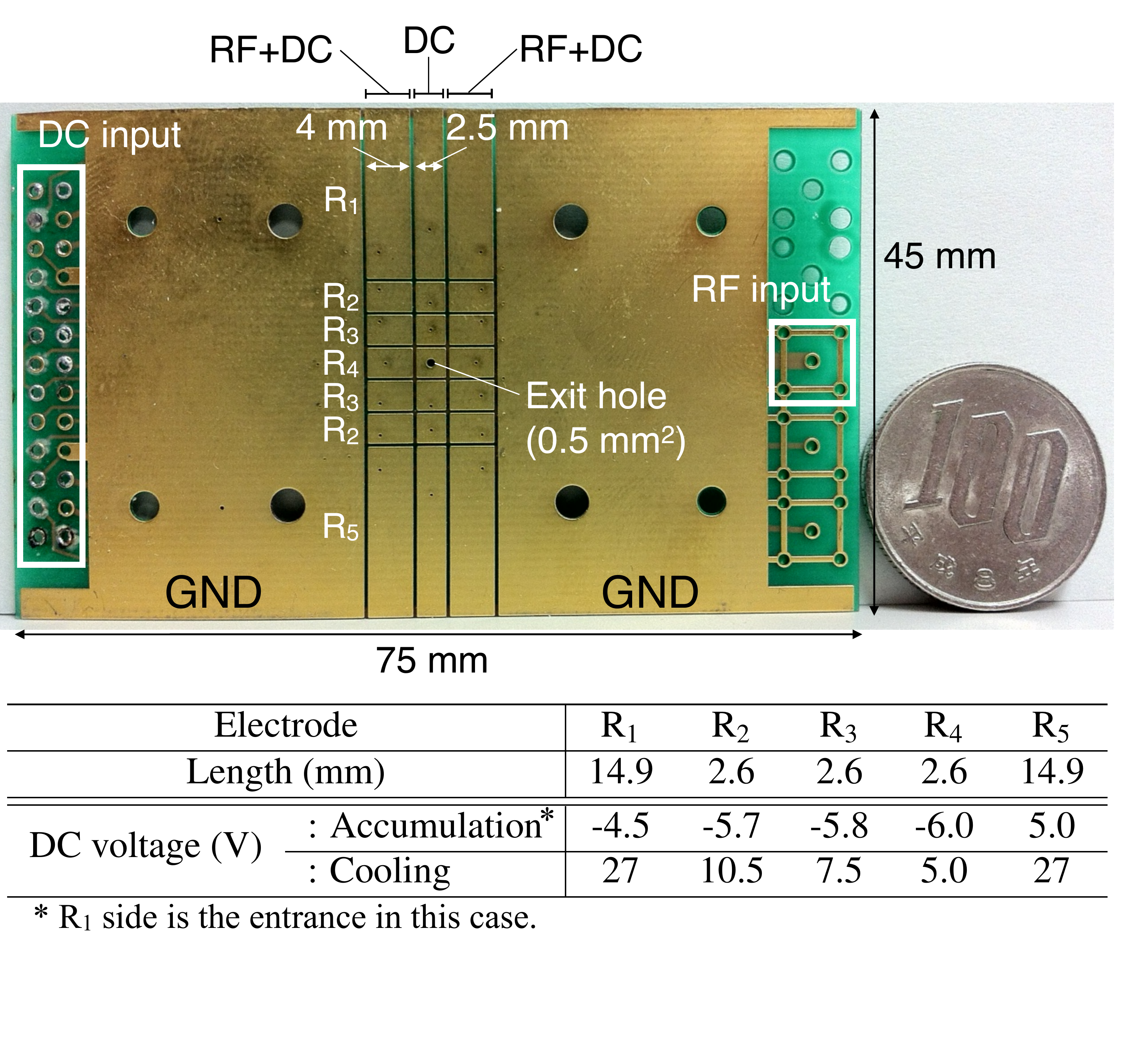} 
 \caption{(color online).
 Photograph of the flat trap PCB with scales and typical DC voltages annotated.
 Capacitors and resistors on the back of the PCB distribute RF and DC signals to the individual electrodes.
 Spacing between adjacent electrodes is 0.3~mm.
 A \yen100 coin is included for scale.
 }
 \label{fig:flat_trap_pattern_001}
\end{figure}

 The flat trap is constructed using two printed circuit boards (PCB) as shown in Fig.~\ref{fig:flat_trap_pattern_001}.
It operates on the same principle as a traditional segmented Paul trap, but uses a novel geometry.  While a traditional Paul trap creates a well-approximated quadrupole field using four rod electrodes, our flat trap design uses six strip electrodes.
While the quadrupole approximation is not sufficient for use as a mass filter, it is perfectly well-suited for ion storage and cooling.
\par The PCBs are mounted on an aluminum block and separated by 4 mm distance.
Each PCB consists of three strips divided into 7 segments (see Fig.~\ref{fig:flat_trap_pattern_001}).
The central electrode of each board has a 0.5 mm$^{2}$ plated hole at its center.
By applying a potential difference between the center electrodes of the PCBs, ions can be extracted orthogonally to the injection axis through the small exit holes.
\par DC electric potentials are used, in the standard manner, to create an axial potential well while an RF signal superimposed on the outer strips provides a radially confining pseudo-potential.
By operating the trap in RF unbalanced mode, there is no need to connect the central strip segments to an RC network, a feature which allows for very fast switching of the centermost segment.
Such fast switching is used to apply a well-defined dipole field at the trap center to eject ions from the trap via the small hole in the central electrode.
This ensures that the ion optical properties of the ion pulse are without higher-order optical aberrations and that the ions have a low emittance.
\section{Results and discussion}
\label{}
\par The performance of the trap system was investigated in terms of the trapping efficiency, cooling time and trap capacity.
Alkali ion sources are capable of providing K and mono-isotopic $^{23}$Na and $^{7}$Li ions.
$^{7}$Li$^{+}$ and $^{23}$Na$^{+}$ were used for the efficiency measurements, while K$^{+}$ ions were used for cooling time and capacity measurements.

\subsection{Trapping in the taper trap}
\label{}
\par In a buffer-gas filled trap, cooled ions don't come out efficiently and quickly without any drag force.
As previously mentioned, a taper structure produces an effective axial drag force.
Due to the drag force, the pre-cooled ions can be efficiently transported to the flat trap.
The axial drag force of the taper trap is determined by the rod angle, $\alpha$.
The value of $\alpha$ was optimized using SIMION.
Simulation showed an increase in transmission with larger angles.
The effective acceptance of the flat trap, however, is estimated to be about 2.8 mm ($\approx$70\% of the PCB gap).
An angle of $\alpha = 2.6$~mrad, corresponding to $\Delta r_{0} = 0.5$~mm for $L = 190$~mm was thus adopted \cite{Ito2010}.

\par
\begin{figure}[ht]
 \centering
 \includegraphics[bb=0 0 947 423, width=86mm]{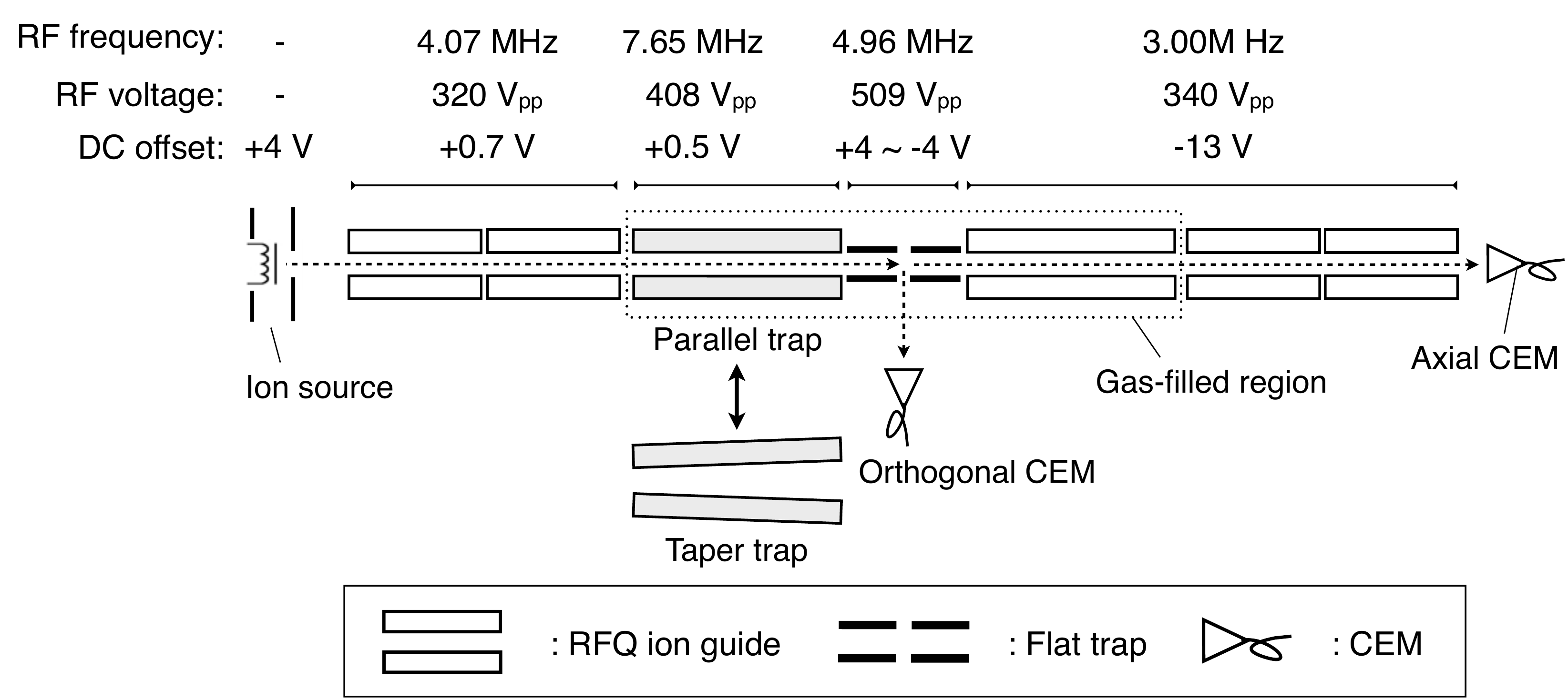} 
 \caption{Experimental setup for the efficiency measurement of parallel and taper trap mode.
 Ions from the alkali ion source were transported by a quadrupole ion guide to the trap system.
 Channeltron electron multiplier (CEM) detectors were installed at the axial and orthogonal exits.
 Typical RF frequencies, RF voltages and DC offset voltages are listed above.
 }
 \label{fig:para_taper_config_001}
\end{figure}

\begin{figure}[ht]
 \centering
 \includegraphics[bb=115 16 908 774, width=86mm]{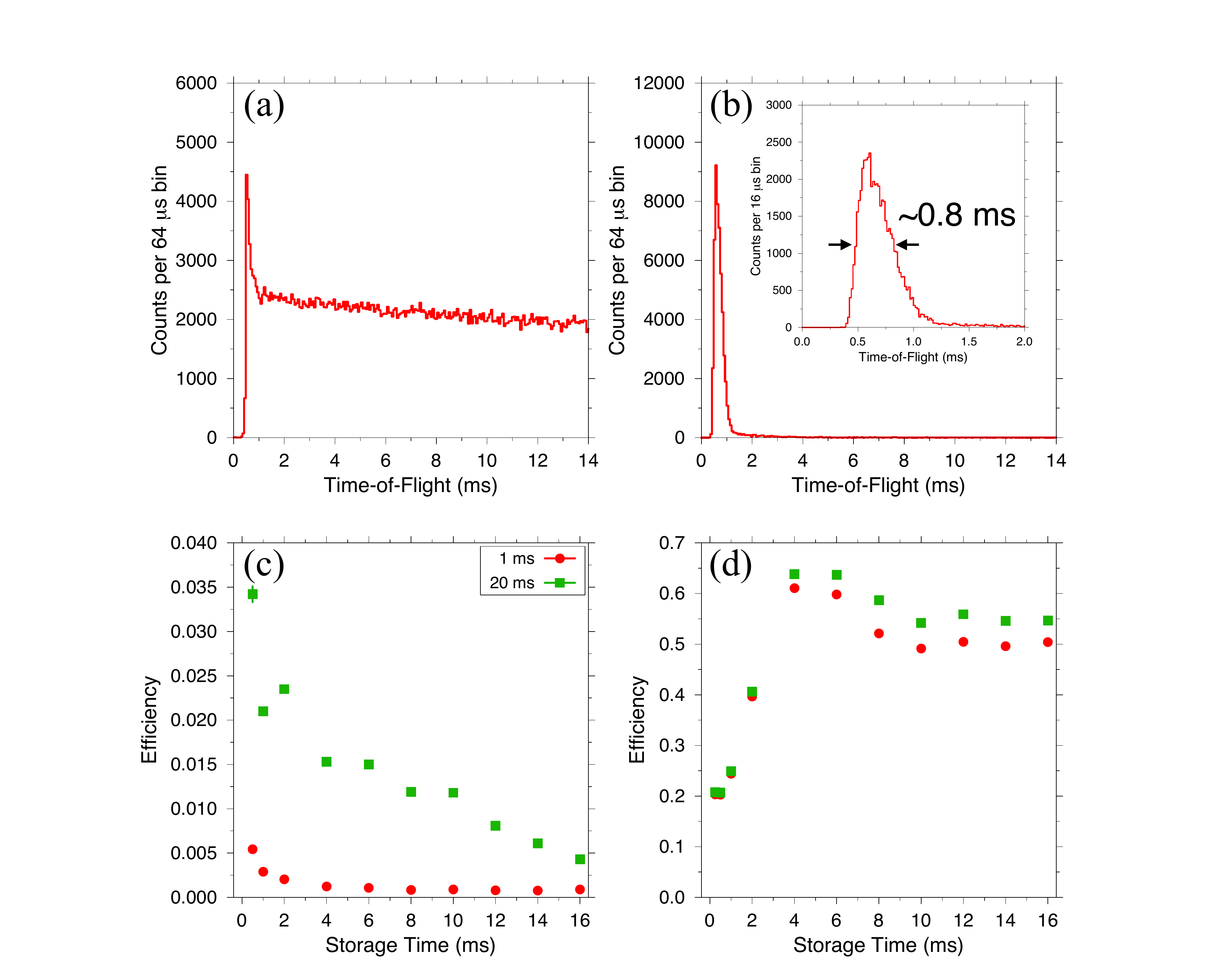}  
 \caption{(color online).
 (top) Typical time-of-flight spectrum measured by CEM at axial exit for (a) the parallel trap, (b) the taper trap. The ion counts are not comparable because the absolute intensity is different in each case.
 (bottom) Transport efficiency as a function of the storage time for (c) the parallel trap, (d) the taper trap.
 For both, the ion counts were integrated over 1 ms (filled squares) and 20 ms (filled circles).
 }
 \label{fig:para_taper_tof_001}
\end{figure}

 Using this geometry, the trapping efficiency and effective drag force of the taper trap were studied.
The experimental setup is shown in Fig.~\ref{fig:para_taper_config_001}.
Due to the taper effective drag, accumulated ions are ejected faster from the taper trap than the parallel trap.
Fig.~\ref{fig:para_taper_tof_001}(a) and (b) show axial ejection time-of-flight spectra for the parallel trap and the taper trap, respectively.
As expected, the spectrum from the parallel trap has a long tail, while the taper trap produces a sharp peak with a width of 0.8~ms.
Nearly 100\% of the ions are extracted within this peak.
This shows that the axial drag force due to the taper structure pushes the ions toward the exit side continuously and such ions can be extracted as a bunch.
\par 
The transport efficiencies from the taper (parallel) trap to the flat trap were measured as follows.
First, the ion rates without trapping in the taper (parallel) trap, $Y_{\rm dc}$, were measured at the axial CEM.
Then, the ion rates after trapping, $Y_{\rm trap}$, were measured at the same detector for various storage times.
The transport efficiencies, $\varepsilon_{\rm trans}$, shown in Fig.~\ref{fig:para_taper_tof_001}(c) and (b) were determined from $\varepsilon_{\rm trans}~=~Y_{\rm trap}/Y_{\rm dc}$
The efficiency of the parallel trap quickly decreases with increasing the storage time, while the taper trap peaks at 4 ms, then reduces to an equilibrium.
For the parallel trap, well-cooled ions, {\it e.g.} after a longer storage time, are hard to extract without an axial drag force, while for the taper trap, the well-cooled ions are efficiently extracted.
This is exemplified by the difference in signals integrated over 1~ms and 20~ms, as shown in the bottom panels of Fig.~\ref{fig:para_taper_tof_001}.

\subsection{Trapping in the flat trap}
\label{}
\par The flat trap is located behind the taper trap.
Its function is to provide final ion bunch preparation for the MRTOF.
The initial size and energy spread of the ion bunch ejected from the flat trap directly affects the efficiency and the mass resolving power of the MRTOF.
The flat trap is required to efficiently provide fast cooling of both axial and radial ion motions, and to eject ions with low emittance.
Ejection is achieved by the novel flat geometry's ability to make a pure dipole field for orthogonal ion ejection.
This is in stark contrast to the axial ejection from a traditional Paul trap, which produces ion pulses with large high-order optical aberrations.
\par Fast cooling is an important feature for the measurement of short-lived nuclei.
In the helium buffer gas of $\sim$$10^{-3}$ mbar at room temperature, the axial cooling is achieved by collisions with He atoms in the DC potential well created by the segmented electrodes, while the radial cooling is achieved by that in the radiofrequency quadrupole pseudo-potential well.
Because of the orthogonal ejection from the flat trap, the axial and radial cooling is partially decoupled; both axial and radial motions must be quickly cooled for effective ion bunching.

\par
\begin{figure}[ht]
  \centering
   \includegraphics[bb=166 20 1149 483, width=86mm]{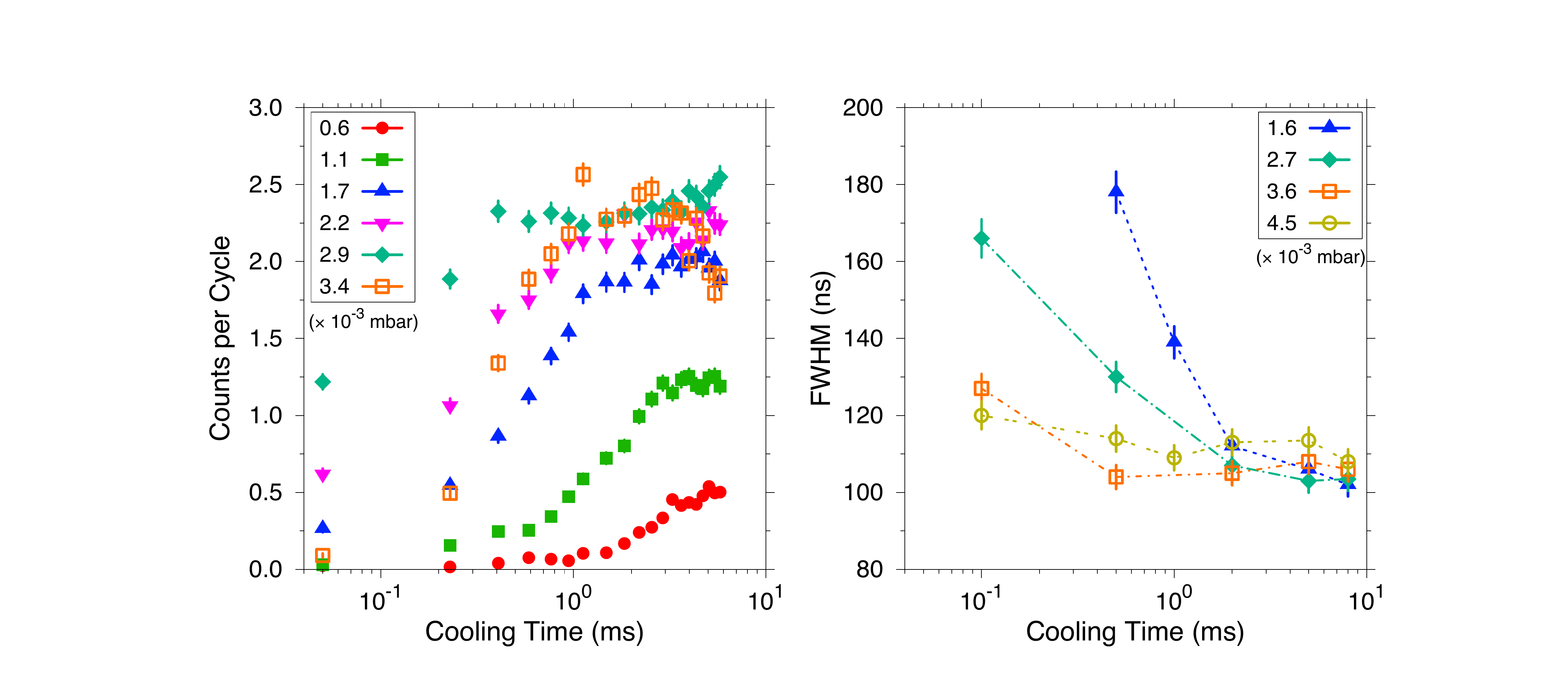}  
  \caption{(color online).
  Two types of cooling time dependences of K$^{+}$ for various gas pressures.
  (left) The ion count rate, measured at the orthogonal CEM, as functions of the cooling time, representing the axial cooling.
  (right) The width of the ToF peak, measured after MRTOF without reflections, as functions of the cooling time, representing the radial cooling.
  }
  \label{fig:cooling_both_001}
\end{figure}

 As the ion cloud cools, it becomes smaller and the velocity of ions decreases.
As the exit hole comprises a much smaller geometric factor along the axis than perpendicular to it, the axial cooling strongly determines the fraction of the ion cloud which can pass through the exit hole.
The radial velocities of the ions, however, contribute to the energy spread and turn-around time, which determines the detected pulse width.
Thus, the rate of detected ions provides a figure of merit for axial cooling, while the pulse width is a figure of merit for radial cooling.
As shown in Fig.~\ref{fig:cooling_both_001}, both the count rate and peak width saturate at around 2~ms with $\approx 3 \times 10^{-3}$ mbar, indicating that the ions accumulated in the flat trap are fully cooled, both axially and radially, within 2~ms.

\subsection{Trapping with double trap}
\label{}


\par For maximum trapping efficiency, the taper trap is used as an auxiliary trap to accumulate and pre-cool an ion bunch while an earlier bunch is being cooled in the flat trap, and to expand the trapping region of the flat trap during the first stage of cooling in the flat trap, thereby, it allows an operational duty cycle of $\approx$100\%.
\par In the operation of 10 ms/cycle, cooling was performed in the flat trap for 3 ms.
At the end of this phase, cooled ions in the flat trap are ejected to the MRTOF.
While pre-cooling was separately performed in the taper trap for 7 ms.
In the following 3 ms period, accumulation was performed with both traps; pre-cooled ions and newly entering ions accumulate in the flat trap.
In the third period, the ions previously accumulated in the flat trap were cooled, just as in the first 3 ms, while new ions, again, enter the taper trap for pre-cooling.
The energy spread of ions entering the flat trap is estimated to be less than ten electron-volts, therefore higher-energy ions may turn back and leave the flat trap.
\par Using the taper trap to initially extend the trapping region, ions can cross over the boundary between the flat and taper trap multiple times until the ion energy is decreased enough to allow them to accumulate in the flat trap.
Due to the effective trap drag force, ions cannot become trapped in the taper region during this ``extended trap" phase.
In the case of the symmetric taper trap, the ion envelope is symmetric near the boundary, while the flat trap entrance is not (see Fig.~\ref{fig:taper_trap_geom_001}(b)).
This can cause some ion loss at the interface with the flat trap during each back and forth pass.
\begin{figure}[ht]
 \centering
 \includegraphics[bb=148 18 916 773, width=86mm]{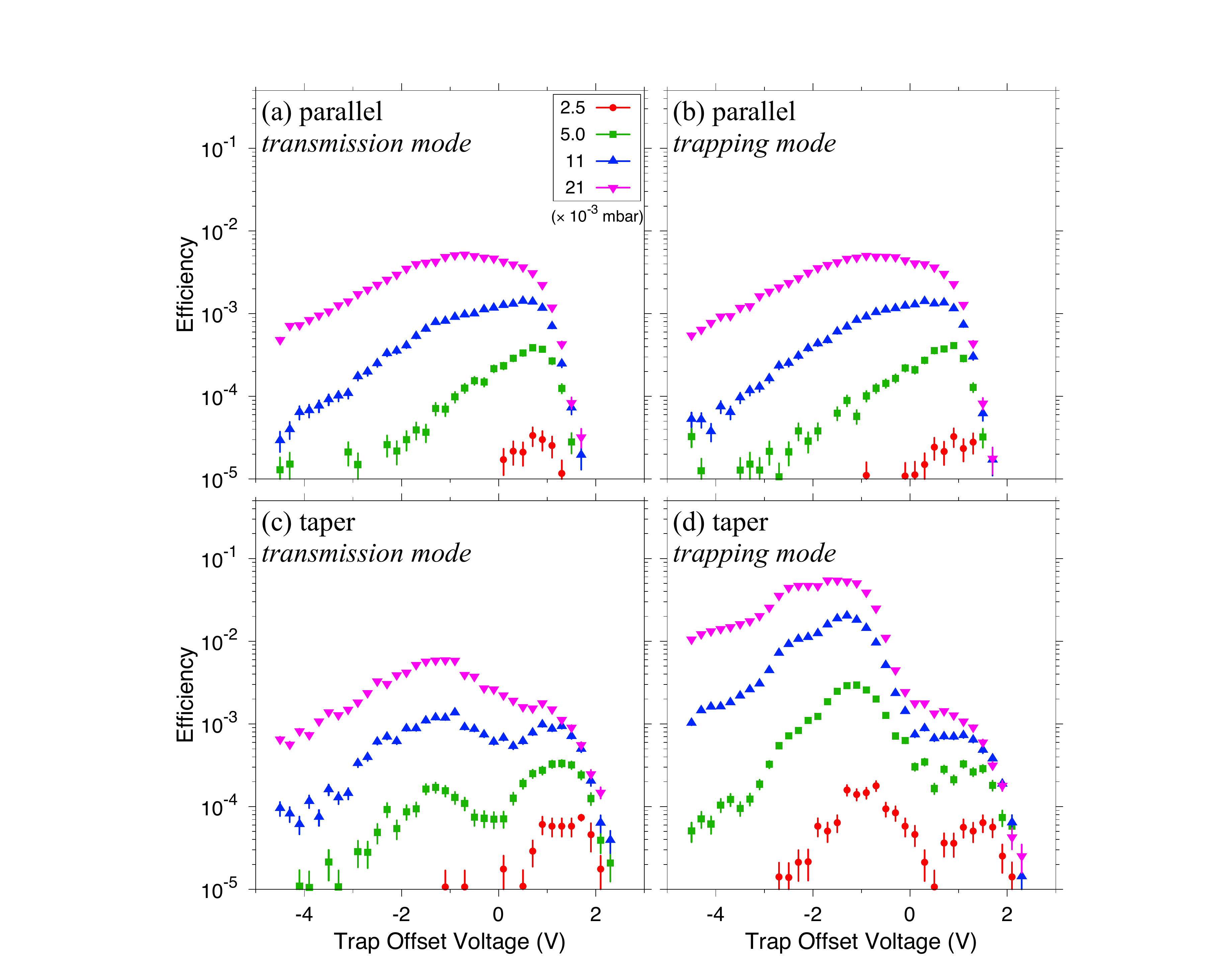}  
 \caption{(color online).
 Trapping efficiency of $^7$Li$^+$ for the parallel trap in (a) transmission mode and (b) trapping mode, and for the taper trap in (c) transmission mode and (d) trapping mode.
 }
 \label{fig:para_taper_efficiency_log_001}
\end{figure}
To avoid this, the inner shape of the taper trap was modified to be asymmetric (see Fig.~\ref{fig:taper_trap_geom_001}(c)).
\par The trapping efficiency using both asymmetric taper (parallel) trap and flat trap, $\varepsilon_{\rm all}$, was measured as follows.
Prior to the trapping efficiency measurement, the ion current was measured at the rods of the taper (parallel) trap and the RFQ ion guides following the flat trap to evaluate the transmission efficiency to the axial CEM.
From the ratio of these, the transmission efficiency of 80(3)\% for DC beam was obtained.
The incoming ion rates, $Y_{\rm dc}$, were measured at the axial CEM and the transmission efficiency was taken into account.
Then, the ion rates ejected from the flat trap, $Y_{\rm trap}$, were measured at the orthogonal CEM for various He gas pressure.
The $\varepsilon_{\rm all}$ shown in Fig.~\ref{fig:para_taper_efficiency_log_001} were determined from $\varepsilon_{\rm all}~=~Y_{\rm trap}/Y_{\rm dc}$.
\par The asymmetric taper (parallel) trap was used with two different operation modes: simple transmission mode and trapping mode.
In the trapping mode, the ions were first accumulated in the taper (parallel) trap for 7 ms to pre-cool prior to accumulation in the flat trap for 3 ms.
While in transmission mode, the ions were simply accumulated in flat trap for 3 ms without pre-cooling.
For the asymmetric taper trap, the $\varepsilon_{\rm all}$ with trapping mode was increased by a factor of ten compared to transmission mode, while for the parallel trap, the $\varepsilon_{\rm all}$ were almost the same for both operation modes.
This indicates again that the pre-cooled ions in the asymmetric taper trap were efficiently transferred to the flat trap due to the drag force.
The trapping efficiencies were determined to be about 27$\%$ for $^{23}$Na$^{+}$ and 5.1$\%$ for $^{7}$Li$^{+}$.

\subsection{Flat trap capacity}
\label{}
\par
\begin{figure}[ht]
 \centering
 \includegraphics[bb=79 6 295 219, width=86mm]{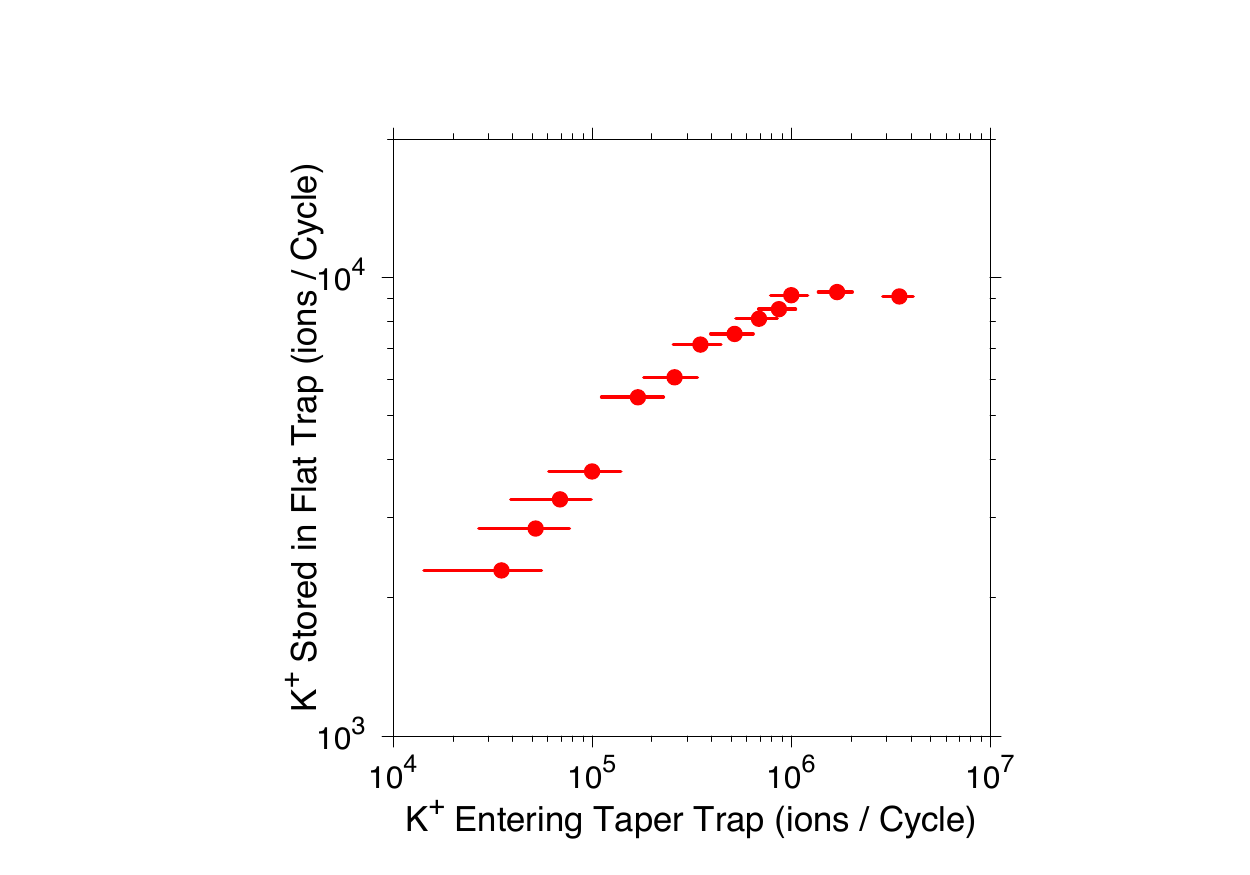}
 \caption{(color online).
 Ion count rate after the MRTOF as a function of the incoming ion rate measured at the rods of the taper trap.
 The K$^{+}$ ion rates were estimated from $^{40}$K$^{+}$ ion rates.
 The statistical error bars for vertical are smaller than symbols.
 }
 \label{fig:trap_capacity_001}
\end{figure}

 In the online experiments, the gas cell may deliver intense molecular impurities.
Too many impurities could saturate the trap.
To estimate the maximum allowable contaminant ratio, the trap capacity was measured using K isotopes with the MRTOF.
The result is shown in Fig.~\ref{fig:trap_capacity_001}.
The intensity of incoming K$^+$ ions was measured at the rods of the taper trap and the total number of trapped K ions was estimated from the $^{40}$K$^{+}$ counts at the microchannel plate detector (MCP) after mass analysis in the MRTOF.
K$^+$ ions were used due to the low abundance (0.0117\%) of $^{40}$K, which allowed for reasonable estimation of the total number of ions in a given pulse.  At high intensities, individual ion signals overlap and ion counting is not possible.  By separating the K isotopes in the MRTOF, the total ion rate could be calculated from the rate of $^{40}$K$^+$ which never saturated due to the very low abundance.  This measurement was performed under different conditions from the measurement of Fig.~\ref{fig:para_taper_efficiency_log_001} and thus the efficiencies cannot be compared.  The flat trap capacity was found to be around $10^{4}$~ions, which is well-consistent with the space charge limit for a few electron-volt potential well.
\section{Conclusions and outlook}
\label{}
\par A novel ion trap system which consists of a flat trap and a taper trap has been developed for ion cooling prior to injection to the MRTOF.
The trapping efficiency, cooling time and trap capacity were investigated.
An efficiency of $\approx$ 27\% for $^{23}$Na$^{+}$ and $\approx$ 5.1\% for $^{7}$Li$^{+}$ were achieved.  Ions were found to fully cool within 2 ms, both axially and radially.
These results satisfy requirements for online experiments using low-yield, short-lived exotic nuclei.  The trap capacity was determined to be $\sim 10^{4}$~ions/cycle, allowing a large contaminant ratio.  Assuming an operation cycle of 100 Hz, incoming ion rates of $\gtrsim$ $10^{6}$ s$^{-1}$ can be acceptable.
\par To improve the performance further, we are planning to mount a cryogenic system, make the conductance smaller and further expand the trapping region with a double taper structure to increase the trapping efficiency and achieve more brilliant ion bunches.
\section{Acknowledgements}
\label{}
\par This work was supported by the Japan Society for the Promotion of Science KAKENHI (Grant Nos.~ 2200823, 24224008 and 24740142).
We also wish to gratefully acknowledge the RIKEN Accelerator Research Facility for their support.





\bibliographystyle{elsarticle-num}

\begin{thebibliography}{99}


\bibitem{Wollnik1990} H. Wollnik and M. Przewloka, Int. J. Mass Spectrom. Ion Proc. 96 (1990) 267
\bibitem{Wada2003} M. Wada, Y. Ishida, T. Nakamura, Y. Yamazaki, T. Kambara, H. Ohyama, Y. Kanai, T. M. Kojima, Y. Nakai, N. Ohshima, A. Yoshida, T. Kubo, Y. Matsuo, Y. Fukuyama, K. Okada, T. Sonoda, S. Ohtani, K. Noda, H. Kawakami and I. Katayama, Nucl. Instrum. Methods Phys. Res., Sect. B 204 (2003) 570
\bibitem{Ishida2005} Y. Ishida, M. Wada and H. Wollnik, Nucl. Instrum. Methods Phys. Res., Sect. B 241 (2005) 983
\bibitem{Schury2009} P. Schury, K. Okada, S. Shchepunov, T. Sonoda, A. Takamine, M. Wada, H. Wollnik and Y. Yamazaki, Eur. Phys. J. A 42 (2009) 343
\bibitem{Schury2013} P. Schury, M. Wada, Y. Ito, S. Naimi, T. Sonoda, H. Mita, A. Takamine, K. Okada, H. Wollnik, S. Chon, H. Haba, D. Kaji, H. Koura, H. Miyatake, K. Morimoto, K. Morita and A. Ozawa, {\it this issue}
\bibitem{Schatz2006} H. Schatz, Int. J. Mass Spectrom. 251 (2006) 293
\bibitem{Takamine2005} A. Takamine, M. Wada, Y. Ishida, T. Nakamura, K. Okada, Y. Yamazaki, T. Kambara, Y. Kanai, T. M. Kojima, Y. Nakai, N. Oshima, A. Yoshida, T. Kubo, S. Ohtani, K. Noda, I.  Katayama, P. Hostain, V. Varentsov and H. Wollnik, Rev. Sci. Instrum. 76 (2005) 103503
\bibitem{Herfurth2001} F. Herfurth, J. Dilling, A. Kellerbauer, G. Bollen, S. Henry, H.-J. Kluge, E. Lamour, D. Lunney, R.B. Moore, C. Scheidenberger, S. Schwarz, G. Sikler and J. Szerypo, Nucl. Instrum. Methods Phys. Res., Sect. A 469 (2001) 254
\bibitem{Mansoori1998} B. A. Mansoori, E. W. Dyer, C. M. Lock, K. Bateman, R. K. Boyd and B. A. Thomson, J. Am. Soc. Mass Spectrom. 9 (1998) 775
\bibitem{SIMION8} SIMION8.0, available at http://www.simion.com
\bibitem{Ito2010} Y. Ito, M. Wada, P. Schury, T. Sonoda, A. Takamine, S. Naimi, H. Wollnik, A. Ozawa and S. Nakamura, RIKEN Accel. Prog. Rep. 44 (2010) 161

\end{thebibliography}



\end{document}